\documentclass[conference]{IEEEtran}
\IEEEoverridecommandlockouts
\usepackage{cite}
\usepackage{amsmath,amssymb,amsfonts}
\usepackage{algorithmic}
\usepackage{graphicx}
\usepackage{textcomp}
\usepackage{xcolor}
\def\BibTeX{{\rm B\kern-.05em{\sc i\kern-.025em b}\kern-.08em
    T\kern-.1667em\lower.7ex\hbox{E}\kern-.125emX}}

\ifCLASSOPTIONcompsoc
  \usepackage[nocompress]{cite}
\else
  \usepackage{cite}
\fi

\usepackage{url}
\usepackage{graphicx} 

\ifCLASSINFOpdf
\else
\fi

\begin{document}
\title{Toward Bundler-Independent Module Federations:\\ Enabling Typed Micro-Frontend Architectures}

\author{\IEEEauthorblockN{Billy Lando}
\IEEEauthorblockA{\textit{University of Bremen}\\
Bremen, Germany\\
Email: billyjov@uni-bremen.de}
\and
\IEEEauthorblockN{Wilhelm Hasselbring}
\IEEEauthorblockA{\textit{Kiel University}\\
Kiel, Germany\\
Email: hasselbring@email.uni-kiel.de}
}

\maketitle

\begin{abstract}
Modern web applications demand scalable and modular architectures, driving the adoption of micro-frontends. This paper introduces Bundler-Independent Module Federation (BIMF) as a New Idea, enabling runtime module loading without relying on traditional bundlers, thereby enhancing flexibility and team collaboration. This paper presents the initial implementation of BIMF, emphasizing benefits such as shared dependency management and modular performance optimization. We address key challenges, including debugging, observability, and performance bottlenecks, and propose solutions such as distributed tracing, server-side rendering, and intelligent prefetching. Future work will focus on evaluating observability tools, improving developer experience, and implementing performance optimizations to fully realize BIMF's potential in micro-frontend architectures.
\end{abstract}


\IEEEpeerreviewmaketitle

\section{Introduction}

In today’s evolving software landscape, developing web applications presents increasing challenges. Applications are becoming larger, development teams are often distributed, and the demand for rapid iteration cycles continues to grow. To manage these challenges, architects have turned to modular approaches like microservices for backend systems and micro-frontends for user interfaces. Microservices divide backend systems into smaller, independently deployable services, while micro-frontends apply similar principles to frontend development. By splitting a large application into smaller, domain-specific micro-frontends, teams can work in parallel, deploy features incrementally, and respond more effectively to changes \cite{Peltonen2021} \cite{antunes2024}.

While micro-frontends offer notable benefits, they also present unique challenges. Managing shared resources—such as libraries, state, or user interface components—without introducing tight coupling or duplication remains a key difficulty. Achieving consistency across independently developed and deployed micro-frontends requires careful coordination \cite{Geers2020, Mezzalira2021}.

Bundlers play a central role in modern web development by helping to address such challenges. A bundler compiles and packages web application assets, such as JavaScript, stylesheets, and images, into production-ready files optimized for deployment. Bundlers also construct a dependency graph to track relationships between source code and third-party libraries, ensuring consistency and reducing errors during development~\cite{webpack5Book}. This process generates one or more bundles, which are optimized files designed to minimize browser requests and improve performance. Bundlers, such as Webpack, are foundational tools for building modern, maintainable web applications \cite{webpack5Book}.

To address the challenges of scaling and modularizing frontend applications, Webpack Version 5 introduced Module Federations \cite{webpack5Book,jackson2020}. This mechanism enables code sharing between independently developed micro-frontends at runtime, bypassing the need for tightly coupled build processes. It facilitates the dynamic sharing of libraries, components, or entire applications, supporting the adoption of scalable, distributed micro-frontend architectures \cite{Mezzalira2021}.

Building on this foundation, Bundler-Independent Module Federation (BIMF) represents a new approach, extending these capabilities by removing the dependency on Webpack and enabling integration with alternative bundlers such as Rspack and Vite.\footnote{\url{https://vite.dev/}} It also supports WebAssembly modules, enabling a broader adoption across diverse development ecosystems. Additional features, including TypeScript integration and enhanced debugging tools, address common issues in large-scale micro-frontend projects, making development workflows more efficient.\footnote{\url{https://module-federation.io/}}

In this paper, we examine the evolution of Module Federations from its introduction in Webpack to its current bundler-independent form. We explore its contributions to micro-frontend architectures, particularly for enhancing modularity, scalability, and deployment flexibility. Finally, we discuss the remaining challenges, such as optimizing performance, improving type safety, and maintaining consistent developer experiences across distributed systems, to identify opportunities for further research and practical impact.

\section{State of the Art}

As web applications continue to grow in scale and complexity, traditional monolithic frontend architectures have increasingly become bottlenecks for large, distributed teams. Micro-frontends, inspired by the microservices paradigm \cite{Pavlenko2020}, address this challenge by decomposing the frontend into independently developed, deployable, and maintainable units. This architectural shift enables teams to work autonomously, adopt tailored workflows, and deliver features faster.

However, this modular approach introduces specific software engineering challenges \cite{Peltonen2021}:

\begin{itemize}
    \item \textbf{Code Duplication and Version Conflicts}: Sharing common libraries and dependencies across independent micro-frontends can lead to duplication, inconsistencies, or version mismatches.
    \item \textbf{Runtime Performance Overhead}: Loading distributed micro-frontends dynamically risks introducing latency and inefficient network usage, particularly when modules have interdependencies.
\end{itemize}
Traditional bundlers like Webpack, Rollup, or Vite were designed to optimize the development and deployment of web applications by compiling and bundling assets into efficient production-ready outputs. While effective for monolithic architectures, these tools were not originally designed to address the specific challenges introduced by micro-frontends.

\subsection{Module Federations in Webpack}

The introduction of Module Federations in Webpack 5 \cite{jackson2020} marked a significant advancement by enabling dynamic module sharing at runtime. Unlike traditional bundling approaches, Module Federations allow micro-frontends to expose and consume shared modules or components without requiring them to be bundled into each application during build time.
Key Benefits of Module Federations are:

\begin{itemize}
    \item \textbf{Dynamic Sharing of Dependencies}: Shared libraries, such as React, are loaded once at runtime, and reused across micro-frontends, reducing redundancy and ensuring consistency.
    \item \textbf{Independent Deployments}: Teams can deploy micro-frontends independently without tightly coupling their build processes.
    \item \textbf{Runtime Decoupling}: Modules are resolved and loaded dynamically based on the host application’s runtime requirements.
\end{itemize}

\begin{figure}[!htb]%
    \centering%
    \includegraphics[width=0.5\textwidth]{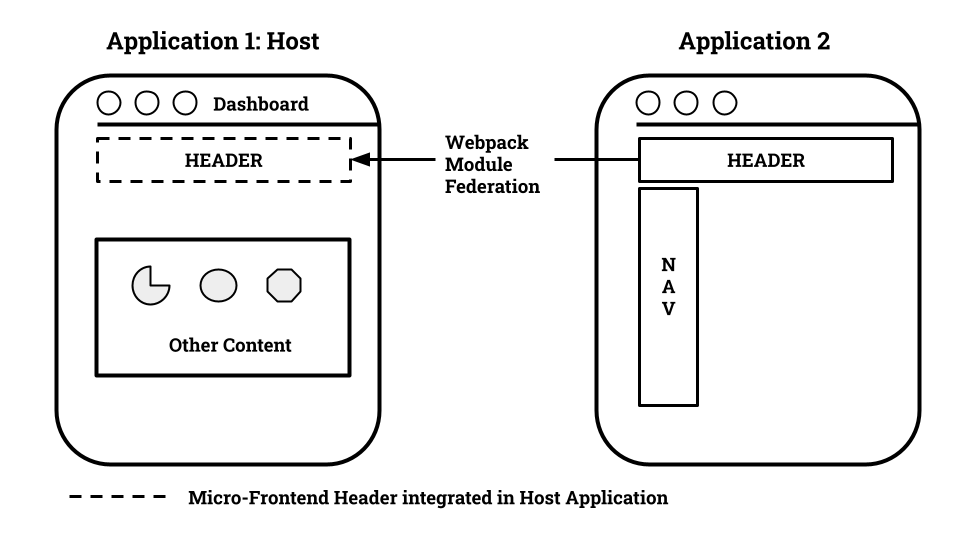}%
    \caption{Header Micro-Frontend Consumption in a Host Application Using Module Federation}%
    \label{fig:mfe}%
\end{figure}

For instance, Figure~\ref{fig:mfe} illustrates a host application, which can dynamically consume a \textit{Header} component from another micro-frontend application, where the \textit{Header} and Navigation components are defined. In this case, the \textit{Header} is exposed as a micro-frontend (illustrated with dashed lines) and loaded into the host application at runtime without being bundled locally. This decoupling minimizes both bundle size and build dependencies while maintaining modularity across applications.

Despite these benefits, Webpack-centric Module Federations pose limitations:

\begin{itemize}
    \item \textbf{Bundler Dependency}: The mechanism is tightly coupled with Webpack’s build system, limiting adoption by teams using other bundlers such as Vite or Rspack.
    \item \textbf{Debugging Complexity}: The distributed nature of Module Federations makes debugging challenging, as it introduces runtime issues like dependency mismatches or load failures.
\end{itemize}

\subsection{The Evolution toward Bundler-Independent Module Federations}

To address these limitations, Module Federation 2.0 emerged as a bundler-agnostic solution, extending the benefits of dynamic module sharing beyond Webpack.\footnote{\url{https://module-federation.io/}} BIMF abstracts the runtime module resolution mechanism, enabling integration with multiple build tools such as Rspack, Vite, Rollup, and ESBuild.

The contributions of Bundler-Independent Module Federations are:

\begin{itemize}
    \item \textbf{Runtime Decoupling}: By removing the dependency on Webpack, BIMF supports diverse bundlers and environments, aligning with modern tool-agnostic workflows.
    \item \textbf{WebAssembly Integration}: WebAssembly (Wasm), a binary instruction format for high-performance execution, allows seamless integration of non-JavaScript languages like Rust into web applications. BIMF leverages this to enhance cross-language compatibility and performance.
    \item  \textbf{Type Safety}: Enhanced TypeScript support ensures type-safe interactions between federated modules, reducing runtime errors caused by incompatible interfaces.
\end{itemize}

The evolution of Module Federations enables greater flexibility and scalability in micro-frontend architectures, accommodating diverse tools, workflows, and team preferences. For example, teams adopting Rspack for its high-performance builds can now leverage Module Federations seamlessly without compromising runtime flexibility.

\subsection{Open Challenges in Micro-Frontend Architectures}

Module Federations have significantly improved micro-frontends, enabling dynamic code sharing at runtime. However, adopting federated architectures still poses several challenges:

\begin{itemize}
\item \textbf{Performance Optimization}: Dynamic loading of federated modules can create request waterfalls, impacting performance. Techniques like Server-Side Rendering (SSR) and data prefetching are under investigation to address this issue.
\item \textbf{Debugging Distributed Systems}: Effective integration of independent micro-frontends requires sophisticated tools for tracing dependencies, diagnosing issues, and monitoring runtime data flows.
\item \textbf{Developer Experience}: The shift to federated architectures poses a steep learning curve, especially with tool-agnostic workflows and runtime module sharing.
\end{itemize}

These challenges highlight the need for ongoing improvements in performance strategies, debugging tools, and developer-friendly workflows tailored to federated micro-frontend systems. The transition from Webpack 5 to Bundler-Independent Module Federation (BIMF) enhances the technology’s versatility, addressing the limitations of traditional bundling and enabling scalable, modular frontend architectures.

\section{The Contributions of Bundler-Independant Module Federation
}

The Bundler-Independent Module Federation (BIMF) extends the concept of Webpack’s Module Federation by decoupling the runtime from specific bundlers, improving debugging, and ensuring type safety. This approach provides greater flexibility, enabling the use of different build tools and supporting cross-language integration, such as using WebAssembly to incorporate modules written in languages like Rust. This section describes the main components of BIMF, how it works, and how it can be applied in micro-frontend web applications.

\subsection{Key Components and Structure of BIMF}

BIMF consists of several key components: 

\begin{itemize} 
\item \textbf{Federated Modules}: These are self-contained pieces of functionality that can be developed, deployed, and loaded dynamically across different micro-frontends. 
\item \textbf{Runtime Abstraction}: BIMF separates the runtime environment from the bundling process, allowing federated modules to work with various bundlers (like Webpack, Vite, or Rspack) without being tied to any one of them. 
\item \textbf{Type Safety}: BIMF integrates TypeScript definitions into the federation process, ensuring that modules follow correct type rules and reducing errors during integration. 
\item \textbf{Debugging Support}: BIMF enhances development tools, such as Chrome DevTools, to make it easier to trace dependencies, monitor module lifecycles, and identify problems. 
\end{itemize}

These components allow BIMF to integrate easily into micro-frontend applications, helping teams manage dependencies, maintain type safety, and simplify debugging, regardless of the bundler used.

\subsection{BIMF in Micro-Frontend Development}

BIMF works by abstracting the runtime from the build process. This means that federated modules can be dynamically loaded at runtime and interact with each other, independent of the bundler used to package them. BIMF supports various bundlers like Webpack, Vite, and Rspack, giving developers the flexibility to choose the most suitable tool for their project.

\begin{itemize} 

\item \textbf{Flexible Tool Selection}: Developers have the flexibility to choose the bundler that best suits the project's requirements, whether it is Rspack for faster builds, Vite for easier configurations, or other bundlers that offer specific advantages. 
\item \textbf{Type Safety}: BIMF ensures modules are correctly typed using TypeScript, which helps prevent errors when integrating modules. \item \textbf{Debugging Support}: BIMF improves debugging tools like Chrome DevTools, enabling developers to inspect shared dependencies, detect performance issues, and fix conflicts as they arise. \end{itemize}

These features help address common challenges in micro-frontend development, such as managing dependencies between modules, debugging issues in a distributed system, and ensuring modules from different teams integrate correctly.

\subsection{BIMF in Micro-Frontend Applications}

In a micro-frontend application, different teams can develop and deploy independent modules that are loaded at runtime. BIMF supports this by allowing:

\begin{itemize} 
 \item \textbf{Independent Deployment}: Each module can be deployed and updated independently, reducing the need for coordination between teams. 
 \item \textbf{Modular Scalability}: New modules can be added to the application without requiring significant changes to the existing codebase. 
 \item \textbf{Reduced Integration Errors}: With type safety and improved debugging tools, BIMF helps prevent issues when integrating modules and speeds up the development process. 
\end{itemize}

These capabilities make BIMF a useful approach for building flexible and scalable web applications with micro-frontend architectures.

\section{Practical Implementation: Use Case Example}

To showcase the practical advantages of Bundler-Independent Module Federation (BIMF), a use case \footnote{\url{https://github.com/billyjov/bimf}} was implemented using Nx workspace\footnote{\url{https://nx.dev/}} with React, Rspack, and Module Federation 2.0.  This implementation showcases how BIMF extends existing concepts of module federation to operate independently of traditional bundlers while leveraging monorepo-based architectures for micro-frontends. 

\subsection{Monorepos and Their Role in Micro-Frontends}

Nx, a leading monorepo tool, simplifies the management of multiple projects in micro-frontend architectures by centralizing them in a single repository. This approach overcomes key challenges of polyrepo setups, such as dependency conflicts, slow feedback loops, and inconsistent testing strategies \cite{10.1145/3328433.3328435}.

Polyrepos, while supporting team autonomy, introduce complications when paired with Module Federation systems. Issues such as version mismatches, duplicated dependencies, and delayed testing are common. Monorepos address these challenges by enabling:

\begin{itemize}
    \item \textbf{Instant Feedback Loops}: Changes in shared code propagate immediately across modules, eliminating the need for versioned releases.
    \item \textbf{Simplified Shared Code Management}: Libraries and components remain within the repository, ensuring consistency and reducing risk when changes are made.
   \item \textbf{Streamlined Dependency Updates}: Centralized management of major dependencies like React to prevent conflicts, while keeping minor dependencies flexible.
\item \textbf{Reliable Integration Testing}: Monorepos facilitate coordinated local services for efficient end-to-end testing, minimizing the need for custom frameworks.
\end{itemize}

\subsection{Use Case Overview}

The use case demonstrates BIMF through two React applications built in an Nx workspace:

\begin{itemize}
    \item \textbf{Host Application}: Serves as the primary micro-frontend, dynamically consuming modules at runtime.
    \item \textbf{Remote Application}: Exposes reusable components, such as a header and navigation bar, for the host application.
\end{itemize}

\subsection{Key Features}
The implementation highlights the following BIMF features:

\begin{itemize}
    \item \textbf{Dynamic Module Loading}: Modules are loaded on demand using Module Federation 2.0, reducing bundle size and enhancing runtime flexibility.
    \item \textbf{Shared Dependency Management}: Dependencies like React and React-DOM are shared at runtime to prevent duplication and improve performance.
    \item \textbf{Rspack Integration}: The use of Rspack, a high-performance bundler optimized for WebAssembly, demonstrates BIMF's bundler-agnostic capabilities and accelerates builds.
\end{itemize}

\subsection{Architectural Overview}

\begin{figure}[!htb]%
    \centering%
    \includegraphics[width=0.5\textwidth]{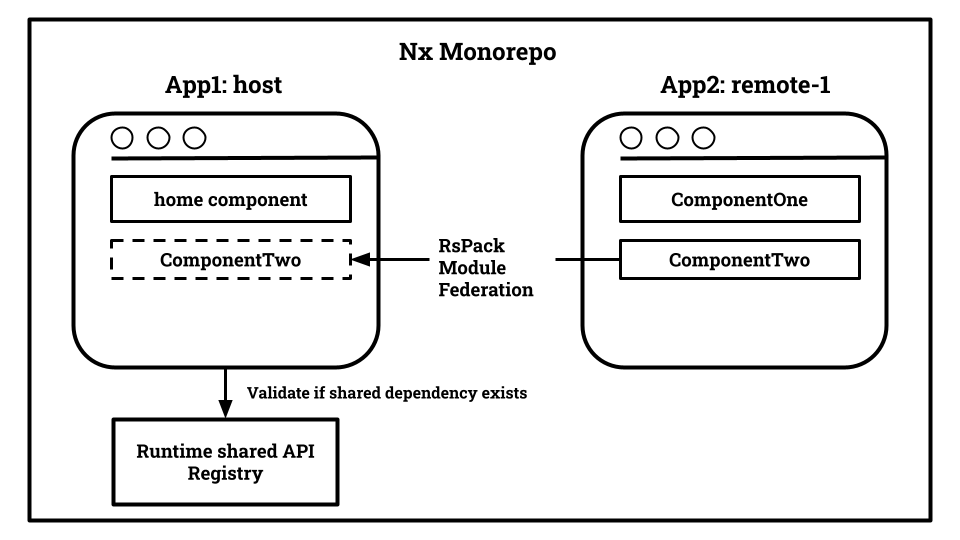}%
    \caption{Consumption of ComponentTwo Micro-Frontend in a Host Application Using Module Federation and Rspack in a Monorepo}%
    \label{fig:bimf}%
\end{figure}

Figure \ref{fig:bimf} shows the architecture, where the host and remote applications function as separate modules within the monorepo. Shared dependencies, such as React and React-DOM, are centrally managed to ensure consistency and improve runtime performance. The runtime shared API registry enables dynamic module loading and integration across various bundlers. While this feature was originally implemented in Webpack, it is now also supported by other bundlers, including Rspack, demonstrating the expanded flexibility of modern Module Federation systems.

\section{Discussion}

The introduction of Bundler-Independent Module Federations presents both opportunities and unresolved challenges.
While it offers a promising approach to modularizing and scaling micro-frontend architectures,
its adoption and practical implementation still face hurdles. This section explores some of
these challenges and outlines open questions for .

\subsection{Debugging and Observability in Distributed Frontend Architectures}

As micro-frontend architectures expand across multiple teams and deployments, debugging and performance
monitoring become complex \cite{Geers2020}. Traditional monolithic systems offer centralized logs and a
clear call stack for troubleshooting. In contrast, federated architectures distribute these responsibilities
across independently deployed modules, often hosted by different teams or even third-party services.
This distribution creates significant challenges in tracing issues, particularly when a problem arises from the
interaction between federated modules.

To address these challenges, it is crucial to establish observability practices that are tailored to federated systems.
Promising approaches include:

\begin{itemize}
\item \textbf{Distributed Tracing Tools}: Adopting tools such as OpenTelemetry,\footnote{\url{https://opentelemetry.io/}} Kieker~\cite{Kieker2020}, Elastic Search \& Kibana,\footnote{\url{https://www.elastic.co/}} and Graphana \& Loki\footnote{\url{https://grafana.com/}} to support seamless collection and correlation of
  logs, metrics, and traces across federated modules.
\item \textbf{Standardized Debugging Protocols}: Developing standards for inter-module communication and error reporting
  to facilitate debugging across team and deployment boundaries.

\end{itemize}

\subsection{Developer Experience and Learning Curve}

Although Bundler-Independent Module Federations introduce powerful new capabilities, their complexity poses a significant barrier to widespread adoption. Developers accustomed
to traditional bundler workflows may struggle to grasp the paradigm shift, particularly the
decoupling of runtime module loading from build tools. Moreover, concepts such as shared
dependencies, dynamic imports, and module versioning require a nuanced understanding
of both build and runtime processes.

Improving the developer experience is essential to drive adoption. Some promising strategies include:

\begin{itemize}
    \item \textbf{Educational Resources}: Providing tutorials, case studies, and visual aids to simplify module federation for developers.
  \item \textbf{Tooling Enhancements}: Enhancing IDEs with live previews, dependency graphs, and auto-configurations for module federation.
\item \textbf{Community Contributions}: Promoting community-shared best practices, reusable setups, and troubleshooting tips.
Future research will evaluate the impact of these strategies on reducing developers' cognitive load.
\end{itemize}

\subsection{Performance Challenges in Federated Architectures}

Performance remains a critical concern in micro-frontend systems, largely due to the "request waterfall problem".
This issue arises when multiple independent modules are loaded sequentially, leading to delays because each module's
loading depends on the previous one's completion. Such delays can negatively impact user experience,
particularly in applications with complex dependencies or when users have slower network connections \cite{performanceBook}.

The developers of Module Federation have recognized this challenge and highlighted plans for future
strategies to address it.\footnote{\url{https://github.com/module-federation/core/discussions/2397}} These strategies include:

\begin{itemize}
    \item \textbf{Server-Side Rendering (SSR)}: Generating parts of the application on the server and delivering them pre-rendered to the user, reducing perceived load time.
  \item \textbf{Data Prefetching}: Anticipating which modules or resources will be needed next and loading them in advance to minimize wait times during interactions.
\end{itemize}

 These strategies show promise but remain in the planning phase and require deeper investigation. Key questions include how to effectively implement SSR in diverse federated systems with independently developed and deployed modules, and what trade-offs arise between prefetching's performance benefits and its additional resource costs, particularly for users on limited devices or networks. Future research will be crucial to fully unlock the performance potential of bundler-independent module federation in micro-frontend architectures.

Addressing these challenges not only influences the success of individual projects, but also shapes the future of modular software design in increasingly decentralized
development environments.

\section{Conclusion and Future Work}

Bundler-Independent Module Federations (BIMF) represents a transformative approach to modularizing micro-frontend architectures by decoupling runtime module loading from build tools. It enables enhanced flexibility, scalability, and collaboration across independently developed components while addressing some of the traditional challenges associated with monolithic or polyrepo-based architectures. Through its tool-agnostic design, BIMF provides new opportunities to adopt modular development practices without being restricted by specific bundlers or build systems.

However, the adoption of BIMF introduces several challenges that require further investigation. Observability remains a critical concern, as debugging and monitoring distributed modules necessitate tailored solutions, such as distributed tracing tools and standardized debugging protocols. In addition, the learning curve for developers must be addressed through educational resources, improved tools, and community-driven best practices to foster broader adoption. Finally, performance optimization, particularly mitigating issues like the "request waterfall problem", requires exploration of strategies such as server-side rendering and intelligent data prefetching, along with careful consideration of trade-offs in real-world applications.

Future work should focus on the effective and efficient implementation of these strategies. Empirical studies that evaluate observability tools and debugging frameworks will provide insights into their scalability and effectiveness in federated architectures. Research into developer-centric tools and educational resources can improve the developer experience, facilitating smoother adoption of BIMF. Moreover, performance-focused investigations into SSR integration and resource-efficient prefetching mechanisms will be crucial for addressing current bottlenecks and optimizing user experiences.
By addressing these challenges, BIMF has the potential to refine the landscape of modular software design, enabling more maintainable, scalable, and high-performance micro-frontend architectures.

\end{document}